\documentclass[prl,twocolumn,showpacs,superscriptaddress]{revtex4-1} 
\usepackage{graphicx}
\usepackage{amsmath}
\usepackage{amssymb}
\usepackage{bm}
\usepackage{hyperref}
\usepackage{multirow}
\usepackage{appendix}
\usepackage{color}
\usepackage{dsfont}

\newcommand{\Z}{\ensuremath{\mathbb Z}}
\newcommand{\abs}[1]{\ensuremath{\left| #1 \right|}}
\newcommand{\nn}{\nonumber}

\begin{document}

\title{$\Z_2$ peak of noise correlations in a quantum spin Hall insulator}
\author{Jonathan M. Edge}
\thanks{These authors contributed equally to this work.}
\affiliation{Instituut-Lorentz, Universiteit Leiden, P.O. Box 9506, 2300 RA Leiden, The Netherlands}
\author{Jian Li}
\thanks{These authors contributed equally to this work.}
\affiliation{D\'epartement de Physique Th\'eorique, Universit\'e de Gen\`eve, CH-1211 Gen\`eve, Switzerland}
\author{Pierre Delplace}
\affiliation{D\'epartement de Physique Th\'eorique, Universit\'e de Gen\`eve, CH-1211 Gen\`eve, Switzerland}
\author{Markus B\"uttiker}
\affiliation{D\'epartement de Physique Th\'eorique, Universit\'e de Gen\`eve, CH-1211 Gen\`eve, Switzerland}
\date{\today}
\begin{abstract}
  We investigate the current noise correlations at a quantum point contact in a quantum spin Hall structure, focusing on the effect of a weak magnetic field in the presence of disorder. For the case of two equally biased terminals we discover a robust peak: the noise correlations vanish at $B = 0 $ and are negative for $B\not = 0 $. We find that the character of this peak is intimately related to the interplay between time reversal symmetry and the helical nature of the edge states and call it the $\Z_2 $ peak.
\end{abstract}
\maketitle

Measurements of current noise correlations can offer remarkable new insights beyond conductance measurements \cite{Blanter2000}. An example of this is the two-particle Aharonov-Bohm effect in which the presence of a flux can only be determined by measuring noise correlations \cite{Samuelsson2004,Neder2007}. 
Quantum spin Hall  (QSH) systems, discovered \cite{Konig2007} after pioneering studies of time reversal invariant band insulators \cite{Kane2005,Bernevig2006}, are no exception in this regard.
So far, various measurements have characterised the QSH effect from several aspects. First the quantised conductance was measured in a QSH bar \cite{Konig2007}. Non-local current measurements subsequently showed that the currents are carried via quantised edge modes \cite{Roth2009}.  The spin polarisation of these edge modes was established very recently \cite{Brune2012}, thereby vindicating the intuitive picture of the QSH effect consisting of two time reversed copies of the quantum Hall effect.
Scanning techniques \cite{Konig2012, Nowack2012, Ma2012} have now provided additional insights into, e.g., inelastic scattering in the QSH systems \cite{vayrynen_helical_2013}. 

To investigate current noise in mesoscopic structures, a central element is the quantum point contact (QPC) \cite{Reznikov1995,Kumar1996}. Theoretical studies of QPCs in QSH systems have shown ways to test the properties of the helical edge states \cite{Hou2009,Teo2009,Dolcetto2012} and determine interaction strengths of the edge modes \cite{Strom2009}. Current noise studies have been performed to distinguish one-and two- particle tunnelling processes at the QPC \cite{Souquet2012}. Correlations between current noises have also been investigated to this end \cite{Lee2012}, as has the effect of interactions on the noise correlations of the current which is backscattered from a QPC \cite{Schmidt2011}. One question remains open, however, despite its direct experimental relevance: how do the noise correlations vary with a magnetic field that breaks the time reversal symmetry (TRS)? In this scenario, the topologically protected edge states are singular at zero field; otherwise disorder becomes crucial. This is the question that we address in the present letter.

\begin{figure}[t]
  \centering
  \includegraphics[width=0.48\textwidth]{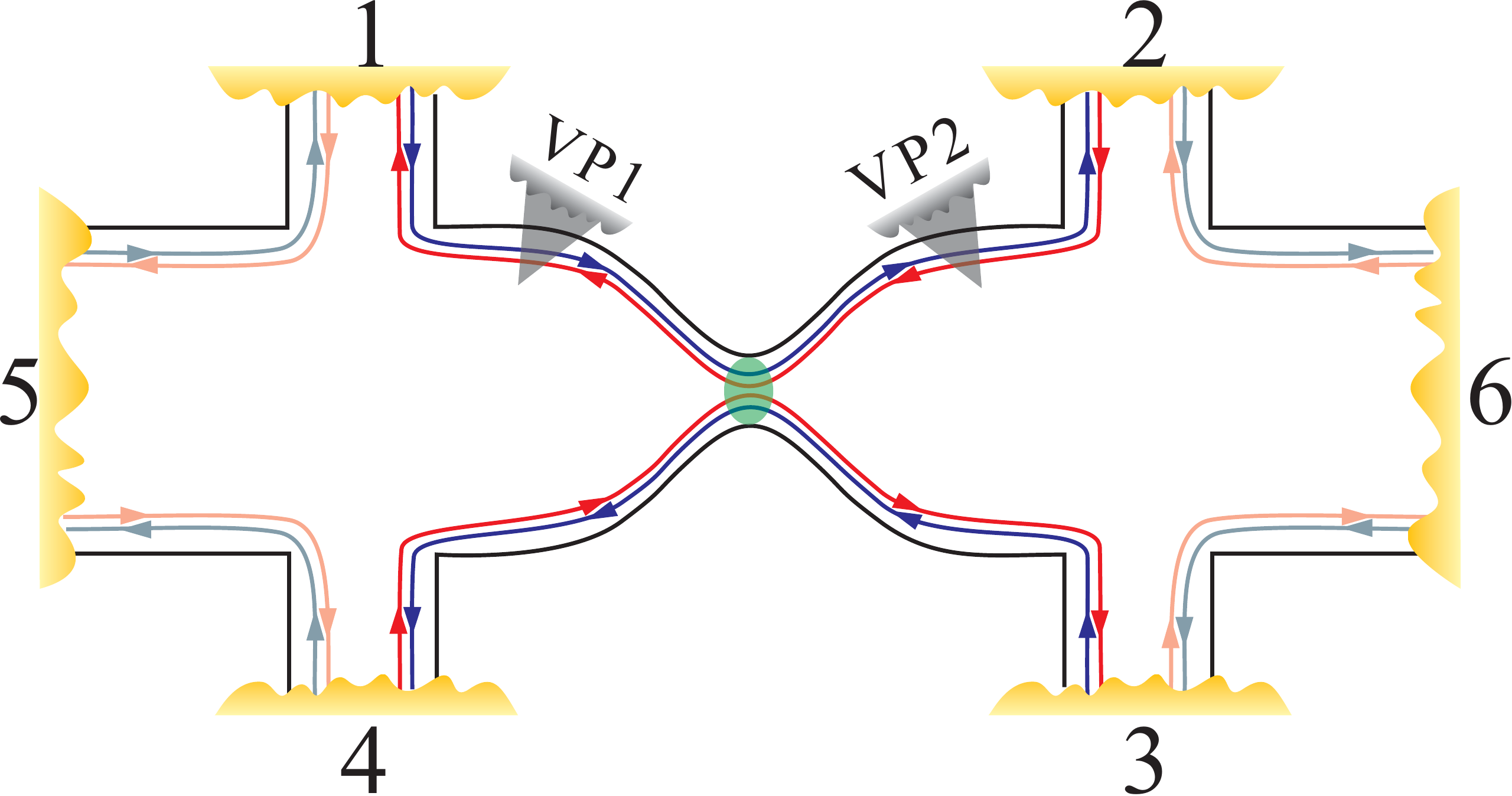}
  \caption{Six terminal Hall bar structure with a quantum point contact. VP 1 and VP 2 are voltage probes which allow for inelastic scattering.
          }
  \label{fig:setup}
\end{figure}

We investigate the noise correlations in a Hall-bar structure with a QPC in a QSH system  (see Fig.~\ref{fig:setup}). Our investigation is based on scattering theory \cite{Buttiker1992}, which assumes the edge modes to be approximately non-interacting channels.
Studies of helical Luttinger liquid theories have shown this to be a good approximation for practical QSH systems in the presence of disorder \cite{Xu2006,Wu2006}, magnetic field \cite{Lezmy2012}, as well as a QPC \cite{Teo2009}.
The relevant scattering matrix in the present setup relating contacts 1 to 4 is a four-by-four matrix (contacts 5 and 6 will be kept grounded). We denote the scattering amplitude for an electron coming from contact $\beta$ and going into contact $\alpha$ to be $S_{\alpha\beta}$. In previous work \cite{Delplace2012}, it has been shown that in the presence of TRS:
$S_{\alpha\alpha}=0,\;S_{13} =-S_{31},\; S_{24}=-S_{42} $ and $S_{\alpha\beta}=S_{\beta\alpha}$ otherwise.
All the off-diagonal entries of $S$ are in general non-zero \cite{Krueckl2011}. An immediate consequence of the vanishing diagonal entries of $S$ is that the equilibrium noise (auto-correlation) at each contact is universal -- it is proportional to the number of open channels connected to the contact \cite{Buttiker1992}, but has no dependence on the details of the QPC. When TRS is broken, the diagonal entries of $S$ become also non-zero, signifying the onset of backscattering, and the matrix $S$ is only subject to unitarity. Using the scattering theory for coherent quantum transport \cite{Buttiker1992}, we can readily write down the noise correlations in terms of the scattering matrix. We will assume in the following the zero-temperature, zero-frequency limit.

We start with the single-source case, namely we set $eV_1 > 0$, $eV_{2,3,4} = 0$, with $V_\alpha$ the voltage at contact $\alpha$. The cross-correlation noise power is then given by \cite{Buttiker1992, Blanter2000}
\begin{align}
  P_{\alpha\beta} = -\frac{e^2}{h} (eV_1) |S_{\alpha 1}|^2 |S_{\beta 1}|^2.
\label{eq:single_source}
\end{align}
This is the partition noise caused by the splitting of the electronic beam at the QPC and it is non-positive.

Next we consider the more interesting case of two biased contacts. To be specific, we set $eV_{1,2} = eV_0 > 0$ and $eV_{3,4} = 0$. We will focus on the current cross-correlations between the two unbiased contacts (3 and 4). In fact the choice of the two contacts to be biased/measured is immaterial.
In this case, the cross-correlation noise power $P_{34}$ contains not only the partition noise similar to Eq.~\eqref{eq:single_source}, but also the exchange noise resulting from scattering of two indistinguishable electrons coming from two different contacts. It is given by
\begin{align}
  P_{34}= - \frac{e^2}h (eV_0) 
  \Bigl[
    &\abs{S_{31}}^2  \abs{S_{41}}^2 + \abs{S_{32}}^2  \abs{S_{42}}^2
  \nn\\
    + &S_{31}^* S_{42}^*  S_{32} S_{41} + S_{31} S_{42} S_{32}^*  S_{41}^*  
  \Bigr]
  .
  \label{eq:p34_1}
\end{align}
The exchange noise, corresponding to the second line of the above equation, can carry nontrivial information encoded in the phases of the scattering amplitudes and manifest it through two-particle interference \cite{splettstoesser_two-particle_2009}. This distinguishes the exchange noise from other measurable quantities to which only scattering \textit{probabilities} are relevant, such as conductance and the pure partition noise.

The total noise power $P_{34}$ is also negative semi-definite, which can be seen by simply rewriting Eq.~\eqref{eq:p34_1} as
$ P_{34}= - (e^3V_0/h) \abs{S_{33}^*S_{43}+S_{34}^*S_{44}}^2
$.
Here the unitarity of the scattering matrix has been used to equate $\abs{S_{31}^*S_{41}+S_{32}^*S_{42}}$ with $\abs{S_{33}^*S_{43}+S_{34}^*S_{44}}$. One important implication of the above equation is: in the presence of TRS, $P_{34}$ reaches its maximum (zero) as $S_{33} = S_{44} = 0$ \footnote{On the other hand if spin-flip scattering process, such as represented by $S_{13}$, were absent, $P_{34}$ would trivially be zero.}; when TRS is broken and backscattering sets in, $P_{34}$ generally becomes negative. We call this peak in the current cross-correlations the $\Z_2$ peak because it is a peculiar phenomenon associated with the form of the scattering matrix of  time-reversal-invariant topological insulators.  It is clear that this phenomenon does not depend on the choices of biased/measured contacts, since we have made no special assumption about the contacts so far.

Physically, the $\Z_2$ peak is a result of an exact cancellation between the partition noise and the exchange noise. It is known \cite{Buttiker1992, Blanter2000} that the partition noise, due to the particle nature of electrons, is negative semi-definite, whereas the exchange noise, due to the fermionic nature of electrons, is positive semi-definite. The two contributions are not necessarily related in generic cases. Here however, TRS and current conservation together demand that the two contributions be of equal magnitude. Similar cancellations can occur in two other circumstances. In one, both outgoing channels are fully occupied at a specific energy. This happens, for example, in a QPC based on chiral edge states \cite{henny_fermionic_1999, oberholzer_hanbury_2000} where both incoming channels are fully occupied at the same energy. In this case the cancellation is trivial because it merely reflects the absence of current fluctuation in each channel. Similarly, a properly-timed mesoscopic two-particle collider with identical sources can lead to a cancellation of the noise correlations \cite{olkhovskaya_shot_2008}, as recently demonstrated in an electronic on-chip experiment \cite{Bocquillon2013}. This case is much closer to the present case, in the sense that the currents in both outgoing channels are noisy by themselves but their correlations vanish identically due to the cancellation.

Remarkably, the $\Z_2$ peak persists even when the incoming channels are subject to strong inelastic scattering. To model this scenario we employ two voltage probes that are coupled to the two incoming arms, from 1 and 2, respectively \cite{buttiker_coherent_1988} (see Fig.~\ref{fig:setup}). For simplicity we assume the same coupling strength $T_p$ for the two voltage probes. $T_p$ is the probability for electrons in the helical channels to enter the additional reservoir connected by a voltage probe. The vanishing total net currents in the voltage probes require the voltage for both additional reservoirs to be
\begin{align}
  V_p = \frac{2+(2-T_p)T+T_p(1-T_p)T^2}{4-(T_pT)^2}V_0\,,
\end{align}
where $T = |S_{21}|^2 = |S_{12}|^2$. In the strong coupling limit, $T_p = 1$ and $V_p = V_0/(2-T)$; the cross-correlation $P_{34}$ measures coherent contributions from the two voltage probe reservoirs instead of the original ones 1 and 2. It is clear that the effect of the voltage probes in this limit is only to substitute $V_0$ in Eq.~\eqref{eq:p34_1} by $V_p$. Such a substitution obviously preserves the qualitative structure of the $\Z_2$ peak.

\begin{figure}[tb]
  \centering
  \includegraphics[width=0.48\textwidth]{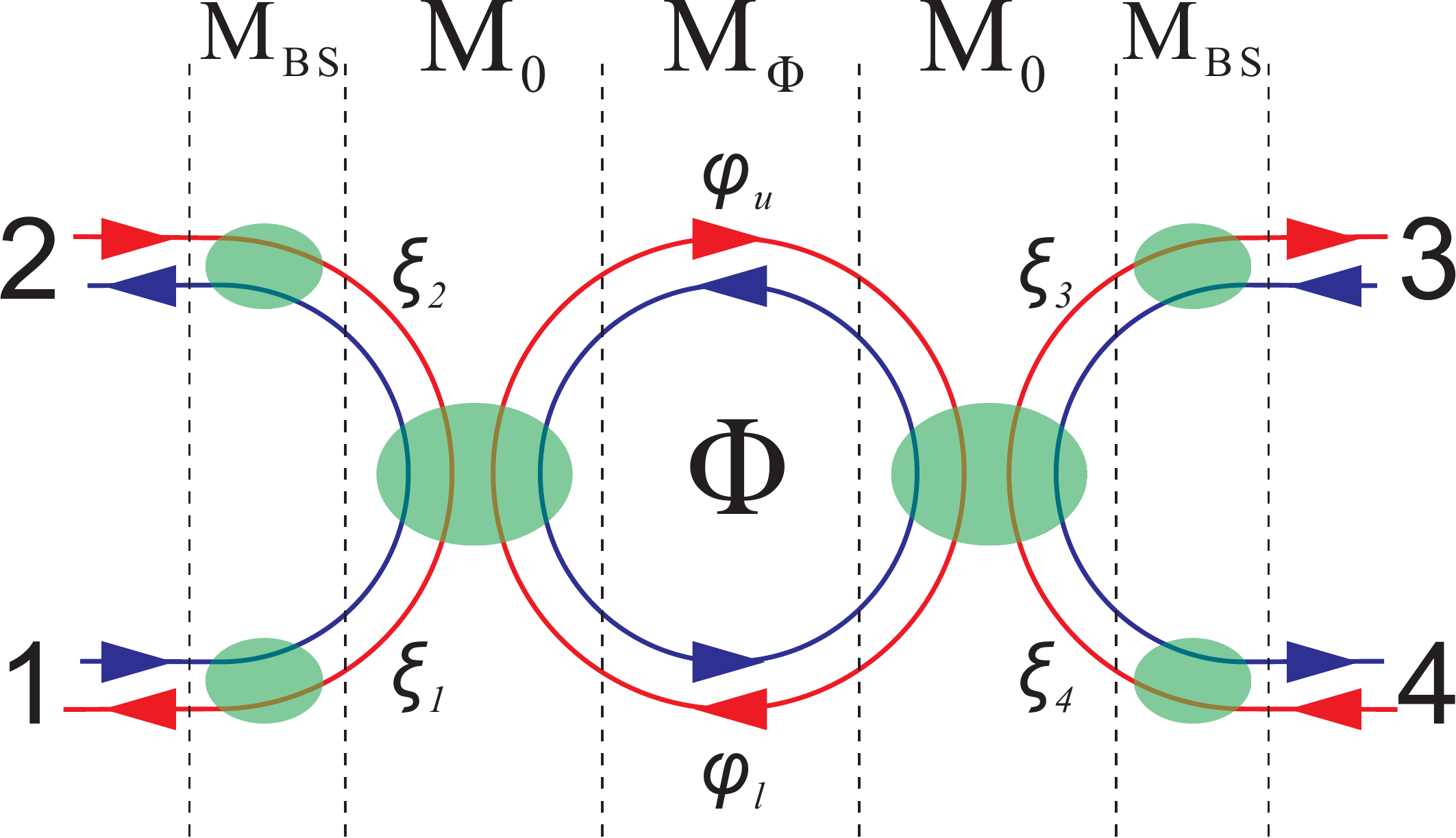}
  \caption{Scattering model for the structure shown in Fig.~\ref{fig:setup} in the presence of a magnetic field and disorder. The setup is partitioned so that the overall scattering matrix can be obtained from combining the transfer matrices (denoted by $M$). A magnetic field causes a flux $\Phi$ in the QPC region and non-zero backscattering along each arm of the helical edge states. Disorder modelled by random scattering phases $\xi_i, \varphi_u$ and $ \varphi_l$  as indicated in the figure leads to a random distribution of the overall scattering matrices.}
  \label{figure2}
\end{figure}

Having established that the suppression of the current cross-correlation is a robust feature of TRS-preserving scattering, we now investigate the effect of a weak magnetic field on the present setup. In real experiments, transport measurements on mesoscopic devices normally display sample-dependent fluctuations when varying the magnetic field, due to disorder \cite{washburn_aharonov-bohm_1986}. In the following we will include disorder, and, as a consequence, investigate the distribution of the cross-correlation noise power $P_{34}$. We consider two major effects of the magnetic field in a generic scenario when disorder is included. The first one is the TRS-breaking scattering at the QPC; the second one is the backscattering that may occur along each arm of the helical edge states before approaching the QPC \cite{Delplace2012}. Here we generally assume that at small magnetic fields the lengths of the paths between leads and the QPC are smaller than the localisation length of the helical edge states \cite{Delplace2012} such that the current cross-correlation is not suppressed simply by localisation. We will model the two effects separately, but consistently with scattering theory.

For the TRS-breaking scattering at the QPC, a minimal model requires an additional loop of helical states inserted into the contact area between the two pairs of original edge states (see Fig.~\ref{figure2}). This loop is coupled to the original edge states in a point-like fashion via TRS tunnelling. Electrons encircling the loop accumulate an Aharonov-Bohm (AB) phase $\Phi$ (cf. Ref~\cite{Delplace2012}). To obtain a scattering matrix for the combined QPC, we adopt the transfer matrix approach.

The transfer matrix describing the local tunnelling between one pair of edge states and the loop states, transformed from a TRS-preserving scattering matrix, is given by
\begin{align}
  M_0 =
  \frac{1}{ \sqrt{1-t^2}}
  \begin{pmatrix}
    t\beta\sigma_x & \beta \\
    \sigma_x\beta\sigma_x & t\sigma_x\beta
  \end{pmatrix},\;
  \beta =
  \frac{1}{\sqrt{1-t^2}}
  \begin{pmatrix}
    -s & r \\
    r & s
  \end{pmatrix},
\end{align}
where $t=|S_{12}|=|S_{34}|$, $s=|S_{13}|=|S_{24}|$ and $r=\sqrt{1-t^2-s^2}=|S_{14}|=|S_{23}|$. $\sigma_{i=0,x,y,z}$ are the conventional Pauli matrices. The transfer matrix for the interior of the loop reads
\begin{align}
  \hspace{-3mm}
  M_{\Phi}
  =
  \begin{pmatrix}
    0 & \gamma(\Phi) \\
    \gamma^*(-\Phi) & 0
  \end{pmatrix},\;
  \gamma(\Phi) =
  \begin{pmatrix}
    0 & e^{i(\Phi+\varphi_l)} \\
    e^{i\varphi_u} & 0
  \end{pmatrix},
\end{align}
where $\varphi_l$ and $\varphi_u$ are the dynamic phases for the lower and upper parts of the loop (see Fig.~\ref{figure2}).
We have chosen a gauge such that the AB phase only enters the lower part of the loop. Transforming the combined transfer matrix $M_{QPC} = M_0M_{\Phi}M_0$, we obtain the scattering matrix for the magnetic-flux-dressed QPC:
\begin{align}
  &S_{QPC} =
  \begin{pmatrix}
    \beta\sigma_x & 0 \\
    0 & \sigma_x\beta
  \end{pmatrix}
  \begin{pmatrix}
    -\Delta_1(\Phi) & \Delta_2(\Phi) \\
    \Delta_2^T(-\Phi) & \Delta_1(\Phi)
  \end{pmatrix}
  \begin{pmatrix}
    \beta\sigma_x & 0 \\
    0 & \sigma_x\beta
  \end{pmatrix},
\label{eq:s_qpc}
\end{align}
with
$\Delta_1(\Phi) = t\sigma_x[\sigma_0+\gamma(\Phi)\gamma(-\Phi)] /[\sigma_0+t^2\gamma(\Phi)\gamma(-\Phi)]$ and \mbox{$\Delta_2(\Phi) = (1-t^2)[\gamma^{\dagger}(-\Phi)+t^2\gamma(\Phi)]^{-1}$}.
In order to illustrate the effect of the magnetic flux on the scattering amplitudes, we extract from Eq.~\eqref{eq:s_qpc} that
\begin{align}
  &S_{33}(\Phi) = S_{44}(\Phi) = \frac{rs}{t} \frac{i\sin\Phi}{\cos\Phi+\cos(\varphi-2i\ln{t})}, \\
  &S_{34}(\Phi) = S_{43}(-\Phi) \nonumber\\
  &\qquad= t\left[1+\frac{s^2}{t^2+e^{-i(\varphi+\Phi)}}+\frac{r^2}{t^2+e^{-i(\varphi-\Phi)}}\right],
\end{align}
where $\varphi = \varphi_l+\varphi_u$. Clearly, the backscattering at the QPC is suppressed when $\Phi=0$ mod $\pi$. On the other hand, by using Eq.~\eqref{eq:p34_1}, we find $P_{34}(\Phi) = P_{34}(-\Phi)$ for the present QPC.

To take into account the backscattering (BS) that occurs along one arm of the helical edge states between a lead and the QPC, we make use of the weak-field-limit result obtained in Ref. \cite{Delplace2012} and write the scattering matrix as
\begin{align}
  S_{BS}(B)=
  \begin{pmatrix}
    -\sqrt{1-\tau(B)^2}e^{i\xi(B)} & \tau(B) \\
    \tau(B) & \sqrt{1-\tau(B)^2}e^{-i\xi(B)}
  \end{pmatrix},
  \label{eq:smat_bs0}
\end{align}
where $\tau(B)=\exp(-\alpha B^2)$ with $\alpha$ being a sample-dependent constant, $\xi(B)$ is a ($B$-dependent) random phase, and $B$ is the magnetic field. The backscattering can be different for different arms, which again relies on specific disorder configurations, but to include all them into the full model is straightforward in terms of the transfer matrix approach (see Fig.~\ref{figure2}). From this we obtain the full scattering matrix.  Details of the above calculation can be found in the supplementary material.

The full scattering matrix contains parameters that are sample-dependent. Varying these parameters allows us to obtain distributions of noise correlations as a function of magnetic field. For simplicity, we fix $t$, $s$, and hence $r$. We also choose a fixed loop area and a fixed $\alpha$ that is the same for all arms. The values of these fixed parameters are determined as described in the supplementary materials and 
they permit a sound comparison with numerical simulations that will be presented below. At a specific $B$, we pick randomly the scattering phases, namely $\varphi_l$, $\varphi_u$ and $\xi$'s, with a uniform probability distribution in $(0,2\pi)$. This turns out to be sufficient to produce a 
random distribution of the full scattering matrix (see below).

\begin{figure}
  \centering
  \includegraphics[width=0.48\textwidth]{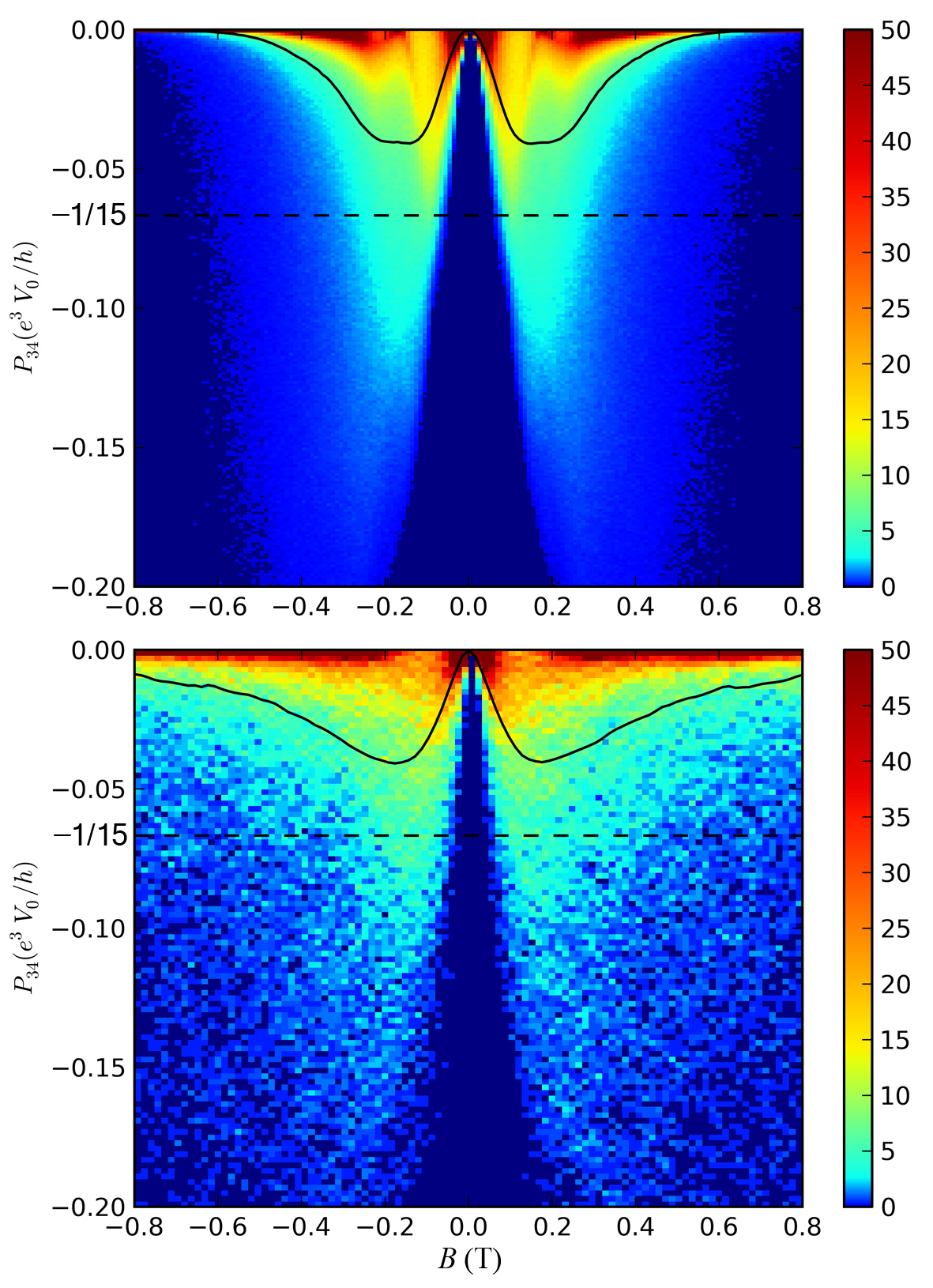}
 \caption{Probability distributions of the cross-correlation noise power $P_{34}$ as a function of magnetic field $B$, obtained from the scattering model with 100,000 disordered samples in the upper panel, and from numerical simulations with 1200 disordered samples in the lower panel. Overlaying both panels, the solid lines show the mean of $P_{34}$, clearly showing the $\Z_2$ peak at $B=0$. The dashed lines indicate the average of $P_{34}$ evaluated for a circular unitary ensemble of scattering matrices.}
  \label{fig:dist}
\end{figure}

The probability distribution of the noise correlation $P_{34}$ produced from the above-described scattering model is plotted in the upper panel of Fig.~\ref{fig:dist}, with the overlaying solid line showing the mean value $\langle P_{34}\rangle$ as a function of $B$. The $\Z_2$ peak of the noise correlation can be clearly identified either from the probability distribution, or more directly in terms of the mean value. The peak structure extends from weak magnetic field up to the point where the noise correlations are suppressed again due to strong backscattering in individual arms. The maximumly negative value of $\langle P_{34}(B) \rangle$ is compared with the average of $P_{34}$ for the circular unitary ensemble of four-by-four scattering matrices, given by the dashed line. The circular unitary ensemble contains uniformly distributed unitary matrices to which the TRS-breaking scattering matrices belong \cite{Beenakker1997}. Averaging Eq.~\eqref{eq:p34_1} in this ensemble yields $\langle P_{34} \rangle_{CUE} = -(1/15)e^3V_0/h$. The maximumly negative value of $\langle P_{34}(B) \rangle$ approaches, but does not reach, $\langle P_{34} \rangle_{CUE}$. This, on the one hand, justifies that by only varying the scattering phases a reasonably random distribution of scattering matrices can be obtained. On the other hand, it also indicates that this random distribution is not quite uniform.

To examine the validity of our scattering model, we further perform numerical simulations with a microscopic Hamiltonian (see supplementary materials for details). We construct numerically a device as illustrated in Fig.~\ref{fig:setup} from a lattice model of the HgTe/CdTe quantum wells \cite{Konig2007}, described at low energy by the Bernevig-Hughes-Zhang (BHZ) Hamiltonian \cite{Bernevig2006}. Disorder is introduced by adding random on-site potentials of a Gaussian profile. The resulting potential fluctuation has a magnitude smaller than the bulk band gap and a correlation length comparable to the penetration depth of the edge states. Scattering matrices connecting transmitting modes between leads are computed from Green's functions \cite{fisher_relation_1981} at various magnetic field for each disorder configuration. The probability distribution of the noise correlation $P_{34}$ is then obtained by using Eq.~\eqref{eq:p34_1}, and plotted in the lower panel of Fig.~\ref{fig:dist}. Comparing with the upper panel, we observe a remarkable agreement between the numerical simulation and the scattering model at weak field, despite a minor quantitative disagreement at stronger field since our choice of $S_{BS}(B)$ in Eq.~\eqref{eq:smat_bs0} is no longer valid.

To summarise, we have constructed a model for a QPC in a QSH system in the presence of disorder, subject to a magnetic field. We have computed the ensemble properties of the noise correlations for this model and found a favourable comparison with results from a numerical calculation. In particular, both approaches show the presence of the $ \Z_2$ peak, a maximum of the noise correlations at zero magnetic field.

We acknowledge useful discussions with C.W.J. Beenakker and T. Schmidt, and helpful comments from Patrick Hofer and Michael Moskalets. This research was supported by NanoCTM, nanoICT, Swiss NSF, MaNEP and QSIT.

\appendix
\begin{widetext}

\section{Supplementary material}

\section{Transfer matrix approach}

We present a derivation of the scattering matrix of figure 2  in the paper (figure~\ref{fig:tmat} in this supplementary material, but with additional labelling) and discuss the number of relevant free parameters.

\begin{figure}
  \centering
  \includegraphics[width=0.75\textwidth]{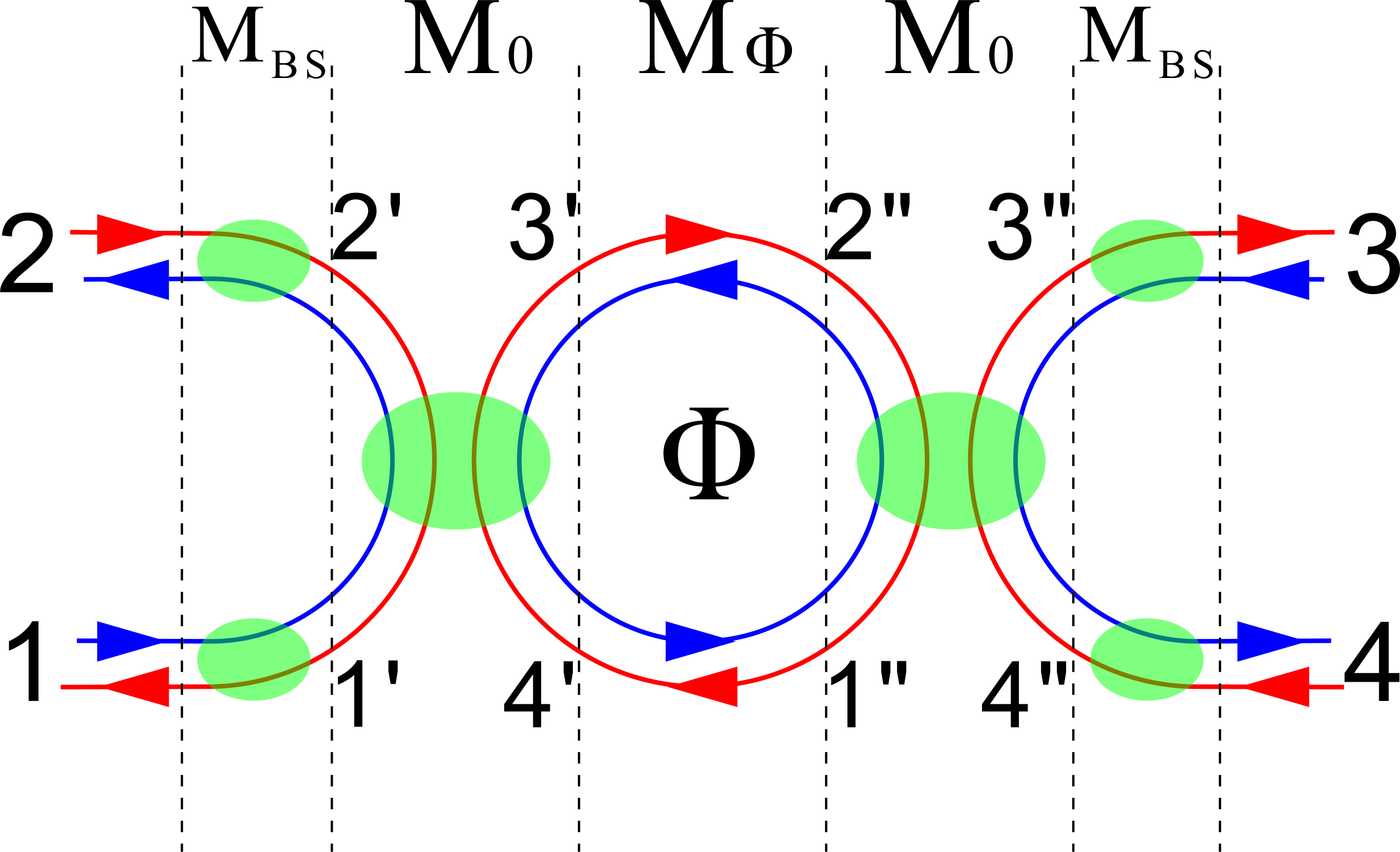}
  \caption{Partitions of the scattering model.}
  \label{fig:tmat}
\end{figure}

\subsection{General transformation and time reversal symmetry}

Define the scattering matrix $S$ as
\begin{align}
  \begin{pmatrix}
    B_L \\
    B_R
  \end{pmatrix}
  = S
  \begin{pmatrix}
    A_L \\
    A_R
  \end{pmatrix}
  =
  \begin{pmatrix}
    S_{LL} & S_{LR} \\
    S_{RL} & S_{RR}
  \end{pmatrix}
  \begin{pmatrix}
    A_L \\
    A_R
  \end{pmatrix},
\end{align}
and transfer matrix $M$ as
\begin{align}
  \begin{pmatrix}
    A_R \\
    B_R
  \end{pmatrix}
  = M
  \begin{pmatrix}
    A_L \\
    B_L
  \end{pmatrix}
  =
  \begin{pmatrix}
    M_{AA} & M_{AB} \\
    M_{BA} & M_{BB}
  \end{pmatrix}
  \begin{pmatrix}
    A_L \\
    B_L
  \end{pmatrix}.
\end{align}
Here, $A$ and $B$ are vectors of current amplitudes (we will always use $A$/$a$ for incoming ones and $B$/$b$ for outgoing ones), $L$ and $R$ stand for left and right sides. The two matrices are related by
\begin{align}
  &M =
  \begin{pmatrix}
    -S_{LR}^{-1}S_{LL} & S_{LR}^{-1} \\
    S_{RL} - S_{RR}S_{LR}^{-1}S_{LL} & S_{RR}S_{LR}^{-1}
  \end{pmatrix}, \\
  \text{or}\quad
  &S =
  \begin{pmatrix}
    -M_{AB}^{-1}M_{AA} & M_{AB}^{-1} \\
    M_{BA} - M_{BB}M_{AB}^{-1}M_{AA} & M_{BB}M_{AB}^{-1}
  \end{pmatrix}.
  \label{M_to_S}
\end{align}

In the system we are considering each block $S_{ij}$, $M_{ij}$ is given by a $2 \times 2$ matrix.
We use the bases defined by
\begin{align}
  \begin{pmatrix}
    b_{1} \\ b_{2}
    \\
    b_{3} \\ b_{4}
  \end{pmatrix}
  =
  S
  \begin{pmatrix}
    a_{1} \\ a_{2}
 \\
    a_{3} \\ a_{4}
  \end{pmatrix},\;
  \begin{pmatrix}
    a_{3} \\ a_{4}
  \\
    b_{3} \\ b_{4}
  \end{pmatrix}
  =
  M
  \begin{pmatrix}
    a_{1} \\ a_{2}
  \\
    b_{1} \\ b_{2}
  \end{pmatrix},
\end{align}
where $a_i/b_i$ denotes an outgoing/incoming channel at terminal $i$.
In this basis time reversal symmetry imposes that the scattering and transfer matrices satisfy the relations
\begin{align}
  S(\mathcal{B}) &= -
  \begin{pmatrix}
    \sigma_z & 0 \\
    0 & \sigma_z
  \end{pmatrix}
  S^T(-\mathcal{B})
  \begin{pmatrix}
    \sigma_z & 0 \\
    0 & \sigma_z
  \end{pmatrix},\\
  M(\mathcal{B}) &=
  \begin{pmatrix}
    0 & -\sigma_z \\
    \sigma_z & 0
  \end{pmatrix}
  M^*(-\mathcal{B})
  \begin{pmatrix}
    0 & \sigma_z \\
    -\sigma_z & 0
  \end{pmatrix},
\end{align}
or
\begin{align}
  S_{LL}(\mathcal{B}) &= - \sigma_z S_{LL}^T(-\mathcal{B}) \sigma_z,\\
  S_{RR}(\mathcal{B}) &= - \sigma_z S_{RR}^T(-\mathcal{B}) \sigma_z,\\
  S_{LR}(\mathcal{B}) &= -\sigma_z S_{RL}^T(-\mathcal{B}) \sigma_z,\\
  M_{AA}(\mathcal{B}) &= \sigma_z M_{BB}^*(-\mathcal{B}) \sigma_z,\\
  M_{AB}(\mathcal{B}) &= -\sigma_z M_{BA}^*(-\mathcal{B}) \sigma_z,
\end{align}
where $\mathcal{B}$ stands for the magnetic field, $L=(1,2)$ and $R=(3,4)$.

\subsection{Single point contact}\label{sec:single_pc}

At a single point contact for helical edge states, the scattering matrix is given by \cite{Delplace2012}
\begin{align}
  \begin{pmatrix}
    b_1 \\ b_2 \\ b_3 \\ b_4
  \end{pmatrix}
  =
  \begin{pmatrix}
    0 & -t & -s & r \\
    -t & 0 & r^* & s^* \\
    s & r^* & 0 & t^* \\
    r & -s^* & t^* & 0
  \end{pmatrix}
  \begin{pmatrix}
    a_1 \\ a_2 \\ a_3 \\ a_4
  \end{pmatrix},
\end{align}
where 1 to 4 label the terminals, and $|t|^2+|s|^2+|r|^2=1$. We can remove the phases of the scattering amplitudes by redefining the current amplitudes (which respects time reversal symmetry) as follows
\begin{align}
  \begin{pmatrix}
    b_1 \\ b_2 \\ b_3 \\ b_4
  \end{pmatrix}
  =
  \begin{pmatrix}
    \xi_{tsr}\tilde{b}_1 \\ \xi_{t}\tilde{b}_2 \\ \xi_{s}\tilde{b}_3 \\ \xi_{r}\tilde{b}_4
  \end{pmatrix},\quad
  \begin{pmatrix}
    a_1 \\ a_2 \\ a_3 \\ a_4
  \end{pmatrix}
  =
  \begin{pmatrix}
   \tilde{a}_1 \\ \xi_{sr}\tilde{a}_2 \\ \xi_{rt}\tilde{a}_3 \\ \xi_{ts}\tilde{a}_4
  \end{pmatrix},
  \label{eq:scattering_phases}
\end{align}
where $\xi_x := x/|x|$. Then the scattering matrix becomes
\begin{align}
  S =
  \begin{pmatrix}
    -t\sigma_x & \sqrt{1-t^2}\beta \\
    \sqrt{1-t^2}\sigma_x\beta\sigma_x & t\sigma_x
  \end{pmatrix},\quad
  \beta = \beta^\dagger = \beta^T = \beta^{-1} =
  \frac{1}{\sqrt{1-t^2}}
  \begin{pmatrix}
    -s & r \\
    r & s
  \end{pmatrix}.
\end{align}
Since we have removed the phases, $r, s, t $ are now real, non-negative numbers.
To test (time reversal) symmetry relations, it will be useful to notice that $\sigma_x\sigma_z\beta\sigma_x\sigma_z=\beta$. 

It is easy to find the transfer matrix, which will be denoted by $M_0$ (see Figure~\ref{fig:tmat}), for the above scattering matrix to be
\begin{align}
  \begin{pmatrix}
    a_3 \\ a_4
  \\
     b_3 \\ b_4
  \end{pmatrix}
  =
  \frac{1}{ \sqrt{1-t^2}}
  \begin{pmatrix}
    t\beta\sigma_x & \beta \\
    \sigma_x\beta\sigma_x & t\sigma_x\beta
  \end{pmatrix}
  \begin{pmatrix}
    a_1 \\ a_2
  \\
      b_1 \\ b_2
  \end{pmatrix}.
\end{align}

\subsection{Point contact including a loop}

The transfer matrix for the loop, which will be denoted by $M_{\Phi}$ (see Figure~\ref{fig:tmat}), is given by
\begin{align}
  \begin{pmatrix}
    a_{1^{''}} \\ a_{2^{''}}
  \\
      b_{1^{''}} \\ b_{2^{''}}
  \end{pmatrix}
  =
  \begin{pmatrix}
    0 & \gamma(\Phi) \\
    \gamma^*(-\Phi) & 0
  \end{pmatrix}
  \begin{pmatrix}
    a_{3^{'}} \\ a_{4^{'}}
   \\
      b_{3^{'}} \\ b_{4^{'}}
  \end{pmatrix},\;
  \gamma(\Phi) =
  \begin{pmatrix}
    0 & e^{i(\Phi+\varphi_l)} \\
    e^{i\varphi_u} & 0
  \end{pmatrix},
\end{align}
where $\varphi_l$ and $\varphi_u$ are the dynamic phases for the lower and upper parts of the loop, $\Phi$ is the AB phase. We have chosen the gauge such that the AB phase only enters the lower part of the loop. Note that the phases appearing in Eq.~\eqref{eq:scattering_phases} and immediately connected to the loop can indeed be absorbed into $\varphi_l$ and $\varphi_u$, therefore the removal of these phases in Sec.~\ref{sec:single_pc} is justified.

The transfer matrix for the full point contact including the loop is given by
\begin{align}
  &\begin{pmatrix}
    a_{3^{''}} \\ a_{4^{''}}
  \\
     b_{3^{''}} \\ b_{4^{''}}
  \end{pmatrix}
  =
  M_{QPC}
  \begin{pmatrix}
    a_{1^{'}} \\ a_{2^{'}}
  \\
      b_{1^{'}} \\ b_{2^{'}}
  \end{pmatrix}
  =
  M_0 M_{\Phi} M_0
  \begin{pmatrix}
    a_{1^{'}} \\ a_{2^{'}}
  \\
      b_{1^{'}} \\ b_{2^{'}}
  \end{pmatrix},\\
  &M_{QPC} =
  \frac{1}{1-t^2}
  \begin{pmatrix}
    \beta\sigma_x & 0 \\
    0 & \sigma_x\beta
  \end{pmatrix}
  \begin{pmatrix}
    \Lambda_1(\Phi) & \Lambda_2^*(-\Phi) \\
    \Lambda_2(\Phi) & \Lambda_1^*(-\Phi)
  \end{pmatrix}
  \begin{pmatrix}
    \beta\sigma_x & 0 \\
    0 & \sigma_x\beta
  \end{pmatrix},\\
  &\Lambda_1(\Phi) = t\sigma_x[\gamma^T(\Phi)+\gamma^*(-\Phi)],\;
  \Lambda_2(\Phi) = \gamma^T(\Phi)+t^2\gamma^*(-\Phi)\,.
\end{align}

The corresponding scattering matrix is given by
\begin{align}
  &S_{QPC} =
  \begin{pmatrix}
    \beta\sigma_x & 0 \\
    0 & \sigma_x\beta
  \end{pmatrix}
  \begin{pmatrix}
    -\Delta_1(\Phi) & \Delta_2(\Phi) \\
    \Delta_2^T(-\Phi) & \Delta_1(\Phi)
  \end{pmatrix}
  \begin{pmatrix}
    \beta\sigma_x & 0 \\
    0 & \sigma_x\beta
  \end{pmatrix},\\
  &\Delta_1(\Phi) = \Delta_1^T(-\Phi) =
  \begin{pmatrix}
    0 & \frac{t(1+e^{i\phi_-})}{t^2+e^{i\phi_-}} \\
    \frac{t(1+e^{i\phi_+})}{1+t^2e^{i\phi_+}} & 0
  \end{pmatrix},\\
  &\Delta_2(\Phi) =
   (1-t^2)
  \begin{pmatrix}
    0 & \frac{e^{-i\varphi_u}}{t^2+e^{i\phi_-}} \\
    \frac{e^{i\varphi_u}}{1+t^2e^{i\phi_+}} & 0
  \end{pmatrix},\\
  &\phi_- = \Phi-\varphi_l-\varphi_u,\; \phi_+ = \Phi+\varphi_l+\varphi_u   \,.
\end{align}

For the lower-right block of $S_{QPC}$, we find
\begin{align}
  &S_{33}(\Phi) = S_{44}(\Phi) = \frac{rs}{t} \frac{i\sin\Phi}{\cos\Phi+\cos(\varphi_l+\varphi_u-2i\ln{t})},\\
  &S_{34}(\Phi) = S_{43}(-\Phi) = t\left[1+\frac{s^2}{t^2+e^{-i(\varphi_l+\varphi_u+\Phi)}}+\frac{r^2}{t^2+e^{-i(\varphi_l+\varphi_u-\Phi)}}\right].
\end{align}
These are the same as equations (7) and (8) in the paper. In a similar way, the other entries of the scattering matrix can be obtained.

\subsection{Including back-scattering}

Unitarity \cite{Beenakker1997} constrains the scattering matrix for back-scattering, using the arm at contact 1 as example, to
\begin{align}
  \begin{pmatrix}
    b_{1} \\
    b_{1'}
  \end{pmatrix}
  &=
  \begin{pmatrix}
    -\rho & \tau\xi_1 \\
    \tau\xi_2 & \rho^*\xi_1\xi_2
  \end{pmatrix}
  \begin{pmatrix}
    a_{1} \\
    a_{1'}
  \end{pmatrix},
  \label{eq:smat_bs0}
\end{align}
where $\rho$, $\tau$, $\xi_1$ and $\xi_2$ are in general all functions of magnetic field $B$; $\tau>0$, $\rho\rho^*+\tau^2=1$, $|\xi_1|=|\xi_2|=1$. The principle of micro-reversibility imposes further constraints: $\rho(-B) = -\rho(B)$, $\tau(B) = \tau(-B)$ and $\xi_2(B)=\xi_1(-B)$. In the following we write $\rho=|\rho|\xi_\rho$ with $\xi_\rho(-B) = -\xi_\rho(B)$.

The transfer matrix corresponding to Eq.~\eqref{eq:smat_bs0} is
\begin{align}
  \begin{pmatrix}
    a_{1'} \\
    b_{1'}
  \end{pmatrix}
  &=
  \frac{1}{\tau}
  \begin{pmatrix}
    |\rho|\xi_\rho\xi_1^* & \xi_1^* \\
    \xi_2 & |\rho|\xi_\rho^*\xi_2
  \end{pmatrix}
  \begin{pmatrix}
    a_{1} \\
    b_{1}
  \end{pmatrix} \\
  &=
  \frac{1}{\tau}
  \begin{pmatrix}
    |\rho|\tilde{\xi}_\rho & 1 \\
    1 & |\rho|\tilde{\xi}^*_\rho
  \end{pmatrix}
  \begin{pmatrix}
    \xi_2 & 0 \\
    0 & \xi_1^*
  \end{pmatrix}
  \begin{pmatrix}
    a_{1} \\
    b_{1}
  \end{pmatrix} \\
  &=
  \frac{1}{\tau}
  \begin{pmatrix}
    \xi_1^* & 0 \\
    0 & \xi_2
  \end{pmatrix}
  \begin{pmatrix}
    |\rho|\xi_\rho & 1 \\
    1 & |\rho|\xi_\rho^*
  \end{pmatrix}
  \begin{pmatrix}
    a_{1} \\
    b_{1}
  \end{pmatrix}, \\
  \tilde{\xi}_\rho &= \xi_\rho\xi_1^*\xi_2^*, \; \tilde{\xi}_\rho(-B) = -\tilde{\xi}_\rho(B).
\end{align}
The above expressions allow us to remove phases $\xi_1$ and $\xi_2$ because changing phases of the final current amplitudes (i.e. $a_i$'s and $b_i$'s) will not lead to any physical effect. Note also that the phases appearing in Eq.~\eqref{eq:scattering_phases} and immediately connected to the back-scattering part can be absorbed into $\xi_\rho$, therefore the removal of these phases in Sec.~\ref{sec:single_pc} is justified.

The back-scattering transfer matrices for the whole setup are then in general
\begin{align}
  &
  \begin{pmatrix}
    a_{1^{'}} \\ a_{2^{'}}
    \\
    b_{1^{'}} \\ b_{2^{'}}
  \end{pmatrix}
  =
  M_{BS}^{(1,2)}
  \begin{pmatrix}
    a_{1} \\ a_{2}
    \\
    b_{1} \\ b_{2}
  \end{pmatrix}
  =
  \begin{pmatrix}
    \frac{\rho_1}{\tau_1} & 0 & \frac{1}{\tau_1} & 0 \\
    0 & \frac{\rho_2}{\tau_2} & 0 & \frac{1}{\tau_2} \\
    \frac{1}{\tau_1} & 0 & \frac{\rho_1^*}{\tau_1} & 0 \\
    0 & \frac{1}{\tau_2} & 0 & \frac{\rho_2^*}{\tau_2}
  \end{pmatrix}
  \begin{pmatrix}
    a_{1} \\ a_{2}
    \\
    b_{1} \\ b_{2}
  \end{pmatrix},\\
  &
  \begin{pmatrix}
    a_{3} \\ a_{4}
    \\
    b_{3} \\ b_{4}
  \end{pmatrix}
  =
  M_{BS}^{(3,4)}
  \begin{pmatrix}
    a_{3^{''}} \\ a_{4^{''}}
    \\
    b_{3^{''}} \\ b_{4^{''}}
  \end{pmatrix}
  =
  \begin{pmatrix}
    \frac{\rho_3}{\tau_3} & 0 & \frac{1}{\tau_3} & 0 \\
    0 & \frac{\rho_4}{\tau_4} & 0 & \frac{1}{\tau_4} \\
    \frac{1}{\tau_3} & 0 & \frac{\rho_3^*}{\tau_3} & 0 \\
    0 & \frac{1}{\tau_4} & 0 & \frac{\rho_4^*}{\tau_4}
  \end{pmatrix}
  \begin{pmatrix}
    a_{3^{''}} \\ a_{4^{''}}
    \\
    b_{3^{''}} \\ b_{4^{''}}
  \end{pmatrix},
\end{align}
where $\tau_i$'s are all real positive and $\rho_i$'s are complex; $\rho_i(-B) = -\rho_i(B)$, $\tau_i(B) = \tau_i(-B)$.

Assuming $\tau_i(B)=\tau(B)=\exp(-\alpha B^2)$ for all $i$, we have
\begin{align}
  &M_{BS}^{(1,2)} =
  \frac{1}{\tau}
  \begin{pmatrix}
    \rho_{(1,2)} & \sigma_0 \\
    \sigma_0 & \rho^*_{(1,2)}
  \end{pmatrix}, \\
  &M_{BS}^{(3,4)} =
  \frac{1}{\tau}
  \begin{pmatrix}
    \rho_{(3,4)} & \sigma_0 \\
    \sigma_0 & \rho^*_{(3,4)}
  \end{pmatrix}, \\
  &\rho_{(1,2)}(B) = -\rho_{(1,2)}(-B) =
  \sqrt{1-\tau^2}
  \begin{pmatrix}
    \xi_{\rho1} & 0 \\
    0 & \xi_{\rho2}
  \end{pmatrix}, \\
  &\rho_{(3,4)}(B) = -\rho_{(3,4)}(-B) =
  \sqrt{1-\tau^2}
  \begin{pmatrix}
    \xi_{\rho3} & 0 \\
    0 & \xi_{\rho4}
  \end{pmatrix}.
\end{align}

Combining everything, we have
\begin{align}
  M_{total} = M_{BS}^{(3,4)}M_{QPC} M_{BS}^{(1,2)}.
\end{align}
From this the total scattering matrix can be obtained using equation~\eqref{M_to_S}.
In total, we have 6 independent random phase factors: $e^{i\varphi_l}$, $e^{i\varphi_u}$ and $\xi_{\rho _i}$ ($i$=1, 2, 3, 4); the first two are $B$-independent, the last four satisfy $\xi_{\rho_i}(-B) = -\xi_{\rho_ i}(B)$.

\section{Numerical simulations}

\begin{figure}
  \centering
  \includegraphics[width=0.75\textwidth]{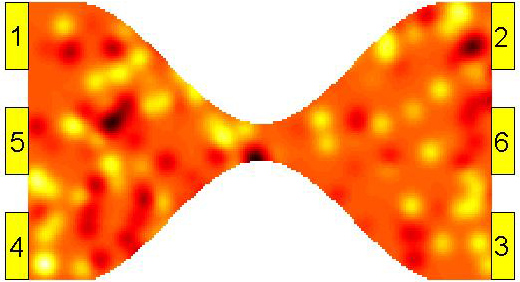}
  \caption{Numerical simulation setup. The total size of the setup is 1$\mu$m$ \times $0.6$\mu$m. The separation of the quantum point contact is 80nm.}
  \label{fig:setup}
\end{figure}

In our numerical simulations, we deal with a setup as shown in Fig.~\ref{fig:setup}. The colors in the interior of Fig.~\ref{fig:setup} encode the electric potential fluctuation caused by disorder. Here, only one specific example of disorder configurations is shown.

\subsection{Hamiltonian}
In the clean limit, we take the Bernevig-Hughes-Zhang (BHZ) Hamiltonian with a spin-orbit coupling term owing to bulk inversion asymmetry \cite{Bernevig2006, Konig2007, konig_quantum_2008}. This Hamiltonian reads
\begin{align}
  H(\bm{k}) = C-Dk^2 + (M-Bk^2)\tau_z\otimes\sigma_0 + Ak_x\tau_x\otimes\sigma_z+Ak_y\tau_y\otimes\sigma_0 + \Delta\tau_y\otimes\sigma_y\,,
\end{align}
where $\tau$'s and $\sigma$'s are Pauli matrices corresponding to orbit and spin degrees of freedom respectively. The parameters are taken from Ref. \cite{konig_quantum_2008}. The orbital effect of the magnetic field is taken into account by substituting $\bm{k}$ with $\bm{k}+(e/\hbar)\bm{\mathcal{A}}$ where $\bm{\mathcal{A}}$ is the vector potential. In numerical simulations, we discretized this Hamiltonian with a lattice spacing $a_0=5$nm.

\subsection{Disorder potential}
The disorder potential, for each impurity center $\bm{r}_i$, is modeled by a Gaussian function:
\begin{align}
  U_i(\bm{r}) = u_i \exp[-(\bm{r}-\bm{r}_i)^2/\lambda^2],
\end{align}
where $u_i$ is randomly taken in a uniform distribution in $(-U_0/2, U_0/2)$, $\lambda$ stands for the range of the potential and is fixed for all impurity centers. We also introduce a parameter $\rho$ for the density of impurity centers. The correlation function for the resultant potential fluctuation is
\begin{align}
  \langle U(\bm{r}) U(\bm{r}+\delta\bm{r}) \rangle = \frac{\pi}{24} U_0^2 (\rho\lambda^2) \exp(-\delta r^2/2\lambda^2)
\end{align}
where $U(\bm{r}) = \sum_i U_i(\bm{r})$. In our simulations, we take $U_0 = 10$meV, $\lambda = 25$nm, and $\rho \lambda^2$ = 0.25.

\subsection{Scattering matrix}
In each normal metal lead connected to the quantum spin Hall insulator sample, there exist a number of transmission modes to ensure a good contact between the lead and the sample. The scattering matrix relates the current amplitudes for these transmission modes. It can be obtained from the Green's functions \cite{mahaux_shell-model_1969, fisher_relation_1981}:
\begin{align}
  S = \mathds{1} - i\sqrt{A} W^\dagger G^R W \sqrt{A}
\end{align}
where $G^R = (E_f+i\eta-H_\text{\scriptsize sample}-W G^R_\text{\scriptsize leads} W^\dagger)^{-1}$ is the retarded Green's function for the sample, $G^R_\text{\scriptsize leads}$ is the retarded Green's function for the semi-infinite leads in their eigen-mode basis, $A = -2\mathrm{Im}(G^R_\text{\scriptsize leads})$ is the spectral function which is diagonal and positive semi-definite, and $W$ is the coupling between the eigen-modes of the leads and the sample. In our simulations we take $E_f = 0.5$meV, where the edge state penetration depth is about 50nm.

\subsection{Comparison with scattering model}

\begin{figure}
  \centering
  \includegraphics[width=0.75\textwidth]{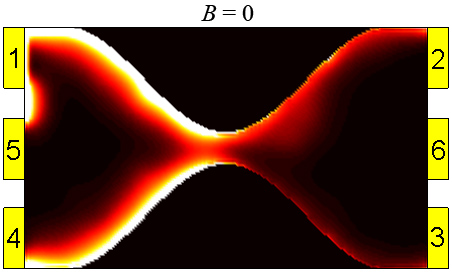}
  \caption{Local density of states as a consequence of biased contact 1.}
  \label{fig:ldos}
\end{figure}

In order to make a sound comparison between the numerical results and the scattering model based on edge states, we determine the fix parameters used in the scattering model in the following way. The area of the loop is estimated from a typical picture of the local density of states around the quantum point contact of our setup (see Fig.~\ref{fig:ldos}). It takes the value $0.012\mu$m$^2$. $t$, $s$ and $r$ are chosen such that, in the absence of magnetic field, the overall scattering probabilities for the QPC obtained by averaging the scattering phases $\varphi_l$ and $\varphi_u$ equal those obtained from numerical simulations. Finally, the factor $\alpha$ is evaluated from a typical dependence of the total transmission probability (from one terminal) on the magnetic field. It takes the value 6T$^{-2}$.










\end{widetext}

\end{document}